\documentclass[a4paper]{jpconf}
\usepackage{graphicx}

\usepackage{color}
\usepackage{xspace}
\usepackage{url}
\usepackage{wrapfig,rotating}
\usepackage{sidecap}
\usepackage{amsmath}
\usepackage{amssymb}
\usepackage{mathrsfs}
\usepackage{lineno}
\usepackage{color}

\newcommand{\ljets}{$\ell +$jets\xspace}
\newcommand{\dzero}     {D0\xspace}

\newcommand{\ttbar}{\ensuremath{t\bar{t}}\xspace}
\newcommand{\mm}       {\mathrm}
\newcommand{\mTT}{\ensuremath{m_{t\bar{t}}}\xspace}

\newcommand{\afb}{\ensuremath{A_{\mbox{{\footnotesize FB}}}^{t\bar{t}}}\xspace}
\newcommand{\afbl}{\ensuremath{A_{\mbox{{\footnotesize FB}}}^{\mm{lep}}}\xspace}

\newcommand{\mcatnlo}   {\textsc{mc@nlo}\xspace}

\begin{document}
\title{Top quark properties at the Tevatron}

\author{Andreas W. Jung}

\address{Fermilab, MS 205, Pine Rd \& Kirk St, Batavia, 60510, IL, USA\\
Preprint: FERMILAB-CONF-14-524-E}

\ead{ajung@fnal.gov}

\begin{abstract}
Recent measurements of top-quark properties at the Tevatron, including top quark production asymmetries and properties, are presented. Latest updates of measurements of top quark production asymmetries include the measurement of the \ttbar production asymmetry by \dzero employing the full Run II data set, in the lepton + jets and dilepton decay channel. Within their uncertainties the results from all these measurements agree with their respective Standard Model expectation.
\end{abstract}

\section{Introduction}
\label{toc:intro}
The top quark is the heaviest known elementary particle and was discovered at the Tevatron $p\bar{p}$ collider in 1995 by the CDF and \dzero collaboration \cite{top_disc1,top_disc2} with a mass around $173~\mathrm{GeV}$. At the Tevatron the production is dominated by the $q\bar{q}$ annihilation process, while at the LHC the gluon-gluon fusion process dominates. The top quark has a very short lifetime, 
\begin{wrapfigure}{r}{0.545\textwidth}
\centerline{\includegraphics[width=0.505\textwidth]{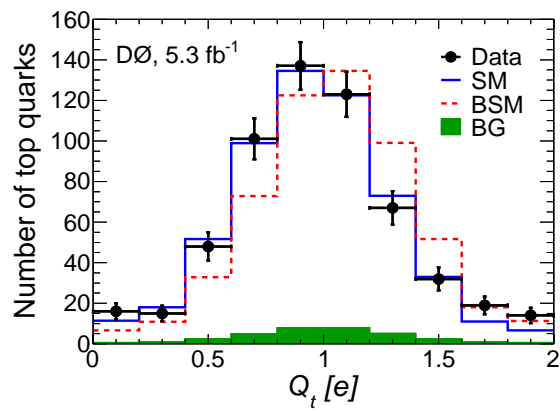}}
\caption{\label{fig:charge} Charge $Q_t$ for \ttbar candidate events in data compared to expectations from the SM and beyond the SM. Background (BG) is indicated by the green-shaded histogram.}
\end{wrapfigure}%
which prevents the hadronization process of the top quark. Instead bare quark properties can be observed.\\
The measurements presented here are performed using either the dilepton ($\ell \ell$) final state or the lepton+jets (\ljets) final state. Within the \ljets~final state one of the $W$ bosons (stemming from the decay of the top quarks) decays leptonically, the other $W$ boson decays hadronically. For the dilepton final state both $W$ bosons decay leptonically. The branching fraction for top quarks decaying into $Wb$ is almost 100\%. Jets originating from a $b$-quarks are identified ($b$-tagged) by means of multi-variate methods employing variables describing the properties of secondary vertices and of tracks with large impact parameters relative to the primary vertex.

%%%%%%%%%%%%%%%%%%%%%%%%%%%%%%%%%%%%%%%%%%%%%%%%%%%%%%%%%%%%%%%%%%%%%%%%
%%%%%%%%%%%%%%%%%%%%%%%%%%%%%%%%%%%%%%%%%%%%%%%%%%%%%%%%%%%%%%%%%%%%%%%%
%%%%%%%%%%%%%%%%%%%%%%%%%%%%%%%%%%%%%%%%%%%%%%%%%%%%%%%%%%%%%%%%%%%%%%%%
% polarization by atlas, single top polarization by cms
\section{Other Top quark properties}
\label{toc:otherProps}
A large variety of measurements of top quark properties at the Tevatron exists to date \cite{d0_web, cdf_web} and is not discussed in detail here. A recent update in terms of top quark properties was done by \dzero. The measurement of the electric charge of top quarks uses \ttbar events in the \ljets channel to fully reconstruct the \ttbar pair and infer the charge of the top quark by employing a jet charge algorithm. Figure \ref{fig:charge} shows the reconstructed top quark charge $Q_t$ distribution compared to the combined MC prediction of signal and background. In addition an exotic model with quarks of $-4/3e$ is shown. Using a data set corresponding to 5.3$/$fb of integrated luminosity the hypothesis that the top quark has a charge of $-4/3e$ is excluded with a significance greater than 5 standard deviations. The results confirm earlier measurements by CDF and ATLAS. An upper limit of 0.46 on the fraction of these exotic quarks in an admixture with SM top quarks is derived at 95\% confidence level.

%%%%%%%%%%%%%%%%%%%%%%%%%%%%%%%%%%%%%%%%%%%%%%%%%%%%%%%%%%%%%%%%%%%%%%%%
%%%%%%%%%%%%%%%%%%%%%%%%%%%%%%%%%%%%%%%%%%%%%%%%%%%%%%%%%%%%%%%%%%%%%%%%
%%%%%%%%%%%%%%%%%%%%%%%%%%%%%%%%%%%%%%%%%%%%%%%%%%%%%%%%%%%%%%%%%%%%%%%%
% afb: -leptonic, ttbar
% intro, atlas 7 tev, cms 7 tev, d0 ljets, dilepton 2D plot, ttbar afb cdf d0, interpretation
\section{Top quark production asymmetries}
\label{toc:angular}
The different initial state makes measurements of angular correlations in $t\bar{t}$ events, such as production asymmetries, complementary between the Tevatron and the LHC. Experimentally, there are two approaches to measure these asymmetries: Either top quarks are fully reconstructed using a kinematic reconstruction or only a final-state particle, e.g. a lepton (`lepton-based asymmetries') is reconstructed. The latter avoids the reconstruction of top-quarks, which is usually more affected by detector resolution and migration effects. The forward-backward asymmetry \afb at the Tevatron measures $\Delta y = y_t - y_{\bar{t}}$, and employing this quantity the production asymmetry is defined as
\begin{equation}
A_{\mbox{{\footnotesize FB}}}^{t\bar{t}} = \dfrac{N(\Delta y >0) - N(\Delta y <0)}{N(\Delta y >0) + N(\Delta y <0)}%,\,\,\mm{and}\,\,A_{\mbox{{\footnotesize C}}}^{t\bar{t}} = \dfrac{N(\Delta |y| >0) - N(\Delta |y| <0)}{N(\Delta |y| >0) + N(\Delta |y| <0)},\,\,\mm{respectively.}
\end{equation}
The lepton-based asymmetries $A_{\mbox{{\footnotesize FB}}}$ and $A^{\ell \ell}$ are defined in the following way employing measurements of the charge $q_{\ell}$ and $\eta$ of the leptons:
\begin{equation}
A_{\mbox{{\footnotesize FB}}}^{\ell} = \dfrac{N(q_{\ell} \cdot \eta >0) - N(q_{\ell} \cdot \eta <0)}{N(q_{\ell} \cdot \eta >0) + N(q_{\ell} \cdot \eta <0)}~,\,\,\mm{and}\,\,A^{\ell \ell} = \dfrac{N(\Delta \eta >0) - N(\Delta \eta <0)}{N(\Delta \eta >0) + N(\Delta \eta <0)},~\mm{respectively}.
\end{equation}
The difference $\Delta \eta$ is given by $\eta_{\ell^+} - \eta_{\ell^-}$ (signs refer to the charge of the lepton). The two lepton-based asymmetries are correlated, but by combining them a small reduction in the uncertainties is obtained.\\
Calculations at NLO QCD including electroweak corrections \cite{bernSi} predicts \afb$=0.088 \pm 0.005$ and \afbl$=0.038 \pm 0.005$. It should be noted that very recently predictions at NNLO+NNLL pQCD by Mitov et al.~became available (discussed at this conference) with a predicted value at NNLO+NNLL including electroweak corrections of \afb$\approx$ 10\%.\\

\begin{figure}[ht]
  \centering
  \includegraphics[width=0.95\columnwidth,angle=0]{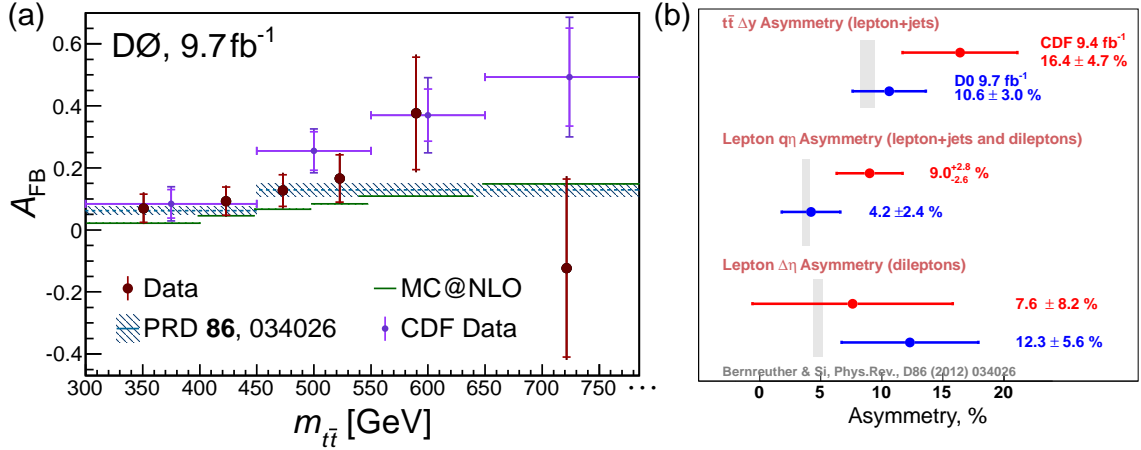}
\protect\caption{\label{fig:cdf_mttafb} The (a) \afb at parton level as a function of the invariant mass of the \ttbar pair $M_{t\bar{t}}$ as measured by CDF and \dzero compared to the predicted dependency by NLO QCD including electroweak corrections \cite{bernSi} or \mcatnlo. Summary of (b) \afb and \afbl measurements at the Tevatron. For \afbl these are results of the combination of results in the \ljets and dilepton decay channel.}
\end{figure}

CDF uses data corresponding to $9.4~\mathrm{fb^{-1}}$ of integrated luminosity and employs a kinematic reconstruction to reconstruct the \ttbar final state in the \ljets decay channel \cite{cdf_ttbar_afb}. CDF measures an inclusive asymmetry of \afb$=0.164 \pm 0.045$ (stat. + syst.) at the parton level compared to the SM prediction of \afb$=0.088 \pm 0.005$ (NLO QCD including electroweak corrections). In addition the kinematic dependency of \afb is extracted, by measuring $\Delta y$ in bins of $M_{t\bar{t}}$, as shown in Fig.~\ref{fig:cdf_mttafb}(a).
% \begin{SCfigure}[][h]
The CDF results show a dependence on $M_{t\bar{t}}$, which is different from the SM expectation by 2.4 standard deviations.\\
\dzero uses the full Run II data, corresponding to $9.7~\mm{fb^{-1}}$ of integrated luminosity \cite{d0_ttbar_afb}, with an improved event selection including the three jet bin, which results in a larger phase space coverage and smaller corrections compared to previous measurements. A likelihood based kinematic reconstruction is used to fully reconstruct the \ttbar final state. The measurement in the \ljets decay channel results in an inclusive asymmetry of \afb$=0.106 \pm 0.030$ (stat. + syst.) at the parton level. The result is compatible with the SM and results by CDF. \dzero does not see an indication for a strong \mTT dependency beyond the one expected by the SM as shown in Fig.~\ref{fig:cdf_mttafb}(a). The recent NNLO+NNLL pQCD calculations are in agreement with the \dzero data.\\

\dzero presented for the first time a measurement of the fully reconstructed top quark asymmetry in the dilepton decay channel. The measurement employs the full Run II data set corresponding to an integrated luminosity of 9.7/fb. Events with at least two jets, two high momentum and isolated electrons or muons or one high momentum isolated electron and muon are selected together with requiring a large missing transverse energy corresponding to the non-detected neutrinos of the leptonic $W$ boson decay. To fully reconstruct the \ttbar event a matrix element technique is applied, which calculates a likelihood of all the possible combinations when assigning reconstructed quantities to parton level \ttbar quantities. 
\begin{figure}[ht]
  \centering
  \includegraphics[width=0.95\columnwidth,angle=0]{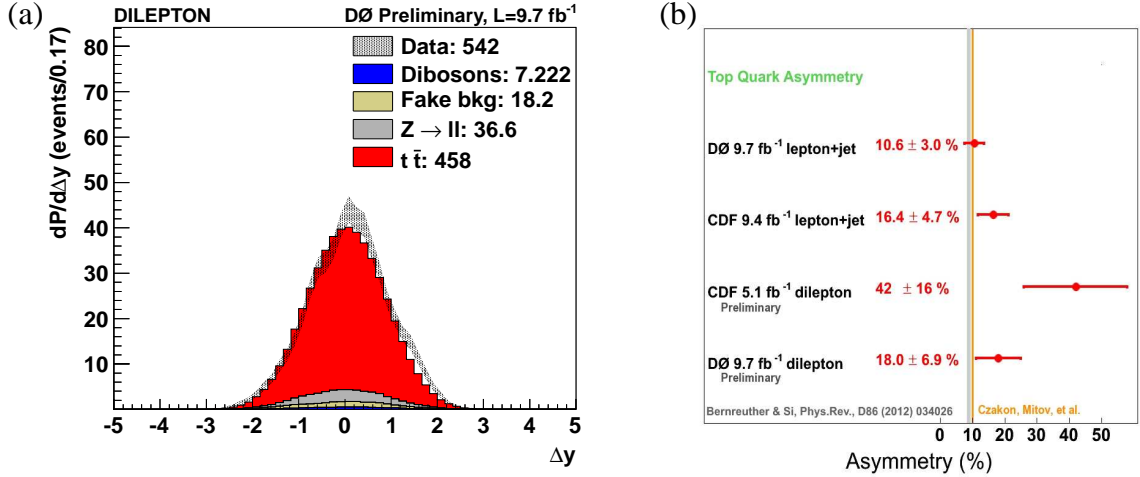}
\protect\caption{\label{fig:d0ll} The (a) $\Delta y$ distribution for selected events compared to the expectation from \mcatnlo and various background contributions. The use of the matrix elements technique reflects in correlated reconstructed $\Delta y$ values, hence data is indicated by the shaded black band. Summary of (b) \afb measurements at the Tevatron compared to predictions at NLO QCD including electroweak corrections \cite{bernSi} and the very recent NNLO+NNLL calculations.}
\end{figure}
Figure \ref{fig:d0ll}(a) shows the $\Delta y = y_t - y_{\bar{t}}$ distribution for the selected data events compared to the signal expectation from \mcatnlo and various background contributions. The measurement is corrected for detector effects to the parton level. If the measurement is interpreted as a test of the SM the measurement yields \afb$=0.180 \pm 0.061\,\mm{(stat.)} \pm 0.032\,\mm{(syst.)}$. Due to the unknown top quark polarization an additional model uncertainty of 5.1\% applies once the measurement is interpreted as a search for contributions of new physics.\\

The \dzero result in terms of the lepton-based asymmetries in the \ljets channel is \mbox{\afbl$= 0.047 \pm 0.026$ (stat. + syst.)} at the parton level \cite{d0_leptonic_afb} and in the dilepton channel the corresponding measurement is \afbl$ = 0.044 \pm 0.039$ (stat. + syst.), whereas the dilepton asymmetry is measured to be $A^{\ell \ell} = 0.123 \pm 0.056$. A summary of \afb and \afbl measurements at the Tevatron is given in Figure \ref{fig:cdf_mttafb}(b). It is interesting to note that the ratio of the two lepton-based asymmetries in the dilepton channel shows a deviation from the SM prediction of about two standard deviations.\\
CDF employed data corresponding to up to $9.4~\mm{fb^{-1}}$ of integrated luminosity and performed a combination of \afbl  measurements. After combining results from \ljets and dilepton channels \afbl is $0.09 ^{+0.028}_{-0.026}$ \cite{CDF-CONF-2013-11035}, see Figure \ref{fig:cdf_mttafb}(b).\\
For measurements of \afb the deviations from the SM predictions got smaller with the new \dzero measurement employing the full data set, but are still higher than the SM predictions. CDF results with the full data set are showing deviations at the two s.d.~level. It should be noted that the individual results on \afb and \afbl employ the full data recorded by CDF and \dzero and Tevatron combinations are currently ongoing.

%%%%%%%%%%%%%%%%%%%%%%%%%%%%%%%%%%%%%%%%%%%%%%%%%%%%%%%%%%%%%%%%%%%%%%%%
%%%%%%%%%%%%%%%%%%%%%%%%%%%%%%%%%%%%%%%%%%%%%%%%%%%%%%%%%%%%%%%%%%%%%%%%
%%%%%%%%%%%%%%%%%%%%%%%%%%%%%%%%%%%%%%%%%%%%%%%%%%%%%%%%%%%%%%%%%%%%%%%%
\section{Conclusions}

Various recent measurements of top quark properties, with a focus on top quark asymmetries, at the Tevatron are discussed. With the wealth of measurements provided by the LHC the measurements at the Tevatron are concentrated on complementary and unique quantities due to the different initial state. In the past results on asymmetry attract quite some interest due to deviations of the measurement compared to SM predictions. \dzero presented the final measurement of \afb in the \ljets decay channel, which is now in good agreement with the theory predictions. CDF remains on the high side. For measurements of \afbl the deviations from the SM predictions got smaller with the new \dzero measurement employing the full data set. Updated asymmetry measurements are forthcoming and studies on combinations of both \afb and \afbl at the Tevatron are currently ongoing. It remains to be seen if the chapter on asymmetry measurements is closed.

\section*{Acknowledgments}
The author thanks the organizers of the TOP 2014 conference for the invitation and for the hospitality of the conference venue.

\end{document}